\documentclass[10pt]{article}
\usepackage{graphicx}

\title{Maximum Cardinality Neighbourly Sets in Quadrilateral Free Graphs}

\author{Neethi K.S.\thanks{Presently with Microsoft, India} {\normalsize{and }}
        Sanjeev Saxena\thanks{E-mail: ssax@cse.iitk.ac.in}\\
Dept. of Computer Science and
Engineering,\\ Indian Institute of Technology,\\
Kanpur, INDIA-208 016}

\date{\today}

\begin{document}
\maketitle

\subsection*{\centering{Abstract}}

Neighbourly set of a graph is a subset of edges which either share an end
point or are joined by an edge of that graph. The maximum cardinality
neighbourly set problem is known to be NP-complete for general graphs.
Mahdian (M.Mahdian, On the computational complexity of strong edge
coloring, {\em{Discrete Applied Mathematics}}, 118:239--248, 2002) proved
that it is in polynomial time for quadrilateral-free graphs and proposed
an $O(n^{11})$ algorithm for the same (along with a note that by a
straightforward but lengthy argument it can be proved to be solvable in
$O(n^5)$ running time). In this paper we propose an $O(n^2)$ time
algorithm for finding a maximum cardinality neighbourly set in a
quadrilateral-free graph.

{\textbf{Keywords:}} Neighbourly Sets, Quadrilateral Free graphs,
$C_4$-free Graphs, Maximum Cardinality Neighbourly Sets, Maximum
Antimatching, Algorithm, Strong Edge Colouring, Antimatching

\section{Introduction}

Two edges $e$ and $e'$ are said to be ``neighbourly'', if and only if,
either $e$ and $e'$ share an end point or if there exists another edge
$e''$ with which both $e$ and $e'$ share a (different) end point. A
neighbourly set $A$ of a graph $G= (V,E)$ is a subset of edges such that
any pair of edges in $A$ are neighbourly. Thus, if $e,e'\in A$, then
either $e$ and $e'$ are incident at a vertex $v\in V$, or $e$ and $e'$
are joined by an edge $e''\in E$.  Cameron[5] introduced neighbourly sets
and discussed some properties of neighbourly sets in chordal graphs.

Mahdian[11], uses the term antimatching for neighbourly sets and showed
that the problem of finding a maximum cardinality neighbourly set is
NP-complete for a general graph. However, for quadrilateral free graphs,
he shows that the problem can be solved in polynomial time and proposed
an $O(n^{11})$ time algorithm. A quadrilateral-free graph (also called a
$C_4$-free graph) is a graph in which there are no quadrilaterals (cycles
of length four).

In this paper we show that a maximum cardinality matching in
quadrilateral free graphs can be found in $O(n^2)$ time. 

We also define ``special neighbourly sets''. A subset $A^*$ of $E$ will
be called a special neighbourly set if for any pair of edges $e,e'\in
A^*$, either $e$ and $e'$ share a common end point or there is an edge
$e''$ of $A^*$ with which $e$ and $e'$ share an end point. Clearly, if
$A^*$ is a special neighbourly set, then it is also a neighbourly set.
The special neighbourly sets, will form a connected graph, while, in
general, neighbourly sets may not be connected. 

Mahdian[11] proves NP-completeness of finding maximum cardinality
neighbourly sets by creating $|V|$ separate sets of $(|E|+1)$ vertices
and joining a different set to each vertex. And then showing that this
graph has a neighbourly set of size $B(|E|+1)$, if and only if, the
original graph has a clique of size $B$. As the neighbourly sets in the
construction are connected, the proof also holds (word by word, without
any change) for special neighbourly sets also. 

Thus the problem of finding a maximum cardinality special neighbourly set
is also NP-complete for general graphs. Our proposed algorithm for
finding a special neighbourly set of maximum cardinality in
quadrilateral-free graphs is simpler (there are fewer cases to consider)
but still takes $O(n^2)$ time. 

A related problem is that of strong edge colouring of graphs. Strong edge
colouring of a graph is a colouring of edges of the graph such that no
two neighbourly edges are given the same colour.  The minimum number of
colours needed to strong-edge colour a graph is called the strong
chromatic index of the graph[9].  In other words, a strong edge-colouring
of a graph $G$ is an assignment of colours to the edges of a graph such
that pair of vertices belonging to different edges with the same colour
are not adjacent. Strong edge colouring in graphs has many applications
like frequency assignment in packet radio networks (see [12,13]).

It may be noted that the strong chromatic index of a graph is at least as
large as the cardinality of the maximum neighbourly set of the graph[11].
Thus, the cardinality of the maximum neighbourly set of a graph (both
special and general) is a lower bound on the strong chromatic index of
the graph. 

The rest of the paper is organised as follows. We first discuss in
Section 2, a related problem of independent interest--- the problem
enumerating all triangles in a quadrilateral free graphs. This algorithm
will be used for finding neighbourly sets. We next discuss in Section 3
some properties of neighbourly sets in quadrilateral-free graphs, and
then use these properties to characterise special and general neighbourly
sets in Sections 4 and 5. We describe algorithms to find maximum
cardinality special and general neighbourly set in Sections 4 and 6. It
is shown in Section 7, that the algorithm for special neighbourly sets
will also give a maximum general neighbourly set in all but one case.
This case is described in Section 6.

\section{Enumeration of Triangles in Quadrilateral-free Graphs}

Alon et.al.[2] and Bezem and van Leeuwen [4] show that all triangles in a
graph $G$ can be found in $O(m\alpha(G))$ time, where arboricity
$\alpha(G)$ of a graph $G=(V,E)$ is the minimum number of forests
required to cover all edges of $E$. Linear time algorithms for
enumeration of triangles are also known for planar graphs [4,7,6].

We next describe a simple algorithm for quadrilateral-free graphs.
Basically, we use the property, that for quadrilateral-free graphs, no
edge of $E$ can be in two triangles (if any edge is in two triangles,
these triangles together will form a quadrilateral).

We first look at each edge in turn and construct the adjacency matrix
$Adj$ of the graph $G$. If an edge $uv$ is present in $E$, then
$Adj[u,v]=Adj[v,u]=1$. Adjacency matrix can be constructed in $O(n+m)$
time without actually initialising all the matrix entries (see for
example, Exercise 2.12 of [1]). Thus, we can then determine whether edge
$uv$ is present in $G$ in $O(1)$ time. Let $v$ be the vertex of minimum
degree. Let $N(v)$ denote the set of neighbours of vertex $v$.

\noindent For each pair of vertices $u,w \in N(v)$ we do the following:\\

/* $u$ and $w$ are neighbours of $v$ */\\

if edge $uw$ is present then\\
begin

/* Presence of edge $uw$ indicates presence of triangle $uvw$. */\\
Output/Process the triangle $\triangle uvw$\\
As each edge can be in at most one triangle, we delete the edges
$uv,vw,wu$ from the graph $G$ and decrement degrees of $u,v$ and $w$
(each) by $1$.

end;\\

\noindent Finally we delete the vertex $v$ and all edges incident at it,
adjusting the degrees of neighbours of $v$.\\ 
And repeat with the next minimum degree vertex (this is the vertex now of
minimum degree).

As we are checking for each pair $(u,w)$ of neighbours of $v$, one step
of the algorithm will take $O({d_{min}}^2)$ time, where $d_{min}$ is the
minimum degree of a vertex in $G$.

As a quadrilateral-free graph contains at most ${n}/{4} (1+\sqrt{4n-3})$
edges (see [10]), the average and hence, the minimum degree $d_{min}$ is
at most $(1+\sqrt{4n-3})/{2}$.  Thus, it takes $O({d_{min}}^2) = O(n)$
time for finding if the minimum degree vertex $v$ is part of a triangle. 

Note that as deletion of an edge or vertex cannot introduce a
quadrilateral, the graph remains quadrilateral-free.  So we can repeat
the step of taking the minimum degree vertex each time, maximum of $n-1$
times. Hence the total time taken is $O(n^2)$.

In case of planar graphs the degree of the minimum degree vertex is at
most five [8]. So each step will take $O(1)$ time, resulting in an $O(n)$
time algorithm for planar quadrilateral-free graphs.

As in quadrilateral-free graph, no edge in $G$ can be in more than one
triangle, the number of triangles can be at most ${m}/{3}=O(n\sqrt{n})$,
thus it may be possible to get a faster algorithm for enumerating
triangles in a quadrilateral free graphs.

\section{Neighbourly Set in Quadrilateral-free graphs} 

In this section we discuss some properties, common to both general and
special neighbourly sets in quadrilateral-free graphs. We will use $E_u$
to refers to the set of edges in the graph incident at vertex $u$. 

{\textbf{Lemma 1:}} If a neighbourly set $A$ of a graph $G=(V,E)$,
contains a triangle, say $\triangle abc$, then $A \subseteq E_a \cup E_b
\cup E_c$, where $E_u$ refers to the set of edges in the graph incident
at $u$.

{\textbf{Proof:}} Let us assume there is an edge $a'b' \not \in E_a \cup
E_b \cup E_c $ in $A$. Since $a'b'$ is not adjacent to any of the edges
of triangle $\triangle abc$ (by assumption), it has to be connected via
an edge to $ab$ (since it is a part of a neighbourly set). Without loss
of generality, let the connecting edge be $aa'\in E$ (Otherwise we can
interchange $a$ and $b$ or $a'$ and $b'$ or both). 

Similarly, edges $bc$ and $a'b'$ must be adjacent to another edge (say)
$xy$. Now there are four possibilities (see Figure 1 below): 

\begin{figure}[h!] 
\centering
\includegraphics[width=3.9in]{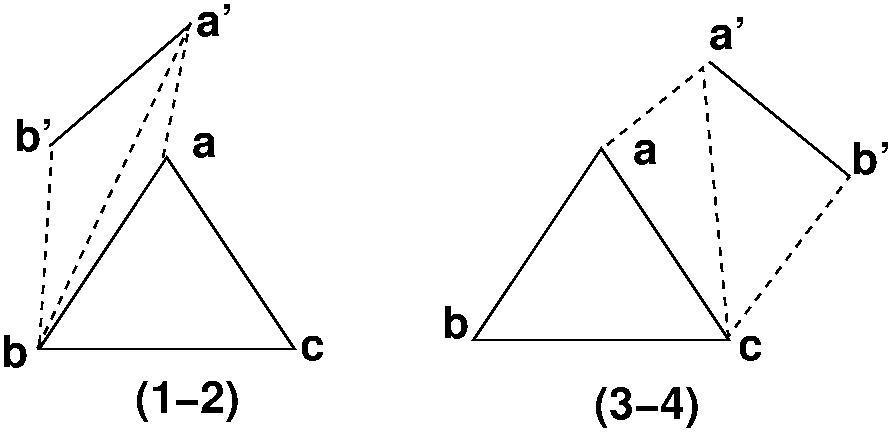}
\caption{Neighbourly set with triangle} 
\end{figure}

{\textbf{Case 1: ($x=b$ and $y=a'$)}} Edge $xy=ba'$. This is not
possible, as $aa'bc$ will form a quadrilateral.

{\textbf{Case 2: ($x=b$ and $y=b'$)}} Edge $xy=bb'$. Not possible as 
$aa'b'b$ forms a quadrilateral.

{\textbf{Case 3: ($x=c$ and $y=a'$)}} Edge $xy=ca'$. Not possible as 
$aa'cb$ is a quadrilateral.

{\textbf{Case 4: ($x=c$ and $y=b'$)}} Edge $xy=cb'$.  Not possible as
$aa'b'c$ is a quadrilateral.

Hence such an edge $xy$ cannot exist and therefore, $a'b'$ cannot be in
$A$. []

For a triangle $abc$, edges in $E_a$ and $E_b$ are both adjacent to $ab$,
thus edges in $E_a \cup E_b \cup E_c$ do form a special neighbourly set.
Therefore, if $A^*$ is a maximal neighbourly set containing a triangle,
then from Lemma 1, $A^* = E_a \cup E_b \cup E_c$ and $A^*$ is also a
special neighbourly set.

{\textbf{Lemma 2:}} If $A$ a neighbourly set of a quadrilateral-free
graph $G=(V,E)$, has a pentagon (i.e., $C_5$, a cycle of length $5$),
then no edge of $A$ can touch any of the pentagonal vertices other than
the five edges forming the pentagon itself.

{\textbf{Proof:}} Let the pentagon be $abcde$. Let us assume there is an
edge, say $p$, distinct from any of the five edges of the pentagon and
touching one of the vertices of the pentagon. There are two possibilities
for this edge (see Figure 2).

{\textbf{$p$ is between any two vertices of the pentagon:}} Without loss
of generality, let one end point of $p$ be $a$.  Now the other end point
cannot be $b$ or $e$, otherwise $p$ will be an edge of the pentagon. So
the edge $p$ can be either $ac$ or $ad$ which will result in presence of
quadrilateral $acde$ or $abcd$ respectively, a contradiction.

{\textbf{$p$ is between one vertex of the pentagon, and the other vertex
is not in pentagon:}} Without loss of generality assume that $p$ is $av$,
where $v$ is the vertex not in the pentagon. 

Since $av$ and $cd$ are both in the neighbourly set there must be an edge
$q$ adjacent to both $av$ and $cd$. From the previous case, we know that
$q$ is neither $ac$ nor $ad$.

Thus, the only remaining possibility is that $q$ is either $vc$ or $vd$. 
If $q$ is $vc$, then we will get the quadrilateral $avcb$ a
contradiction. And if $q$ is $vd$ then we get the quadrilateral $avde$ 
again a contradiction. []

\begin{figure}[h!] 
\centering
\includegraphics[width=4.4in]{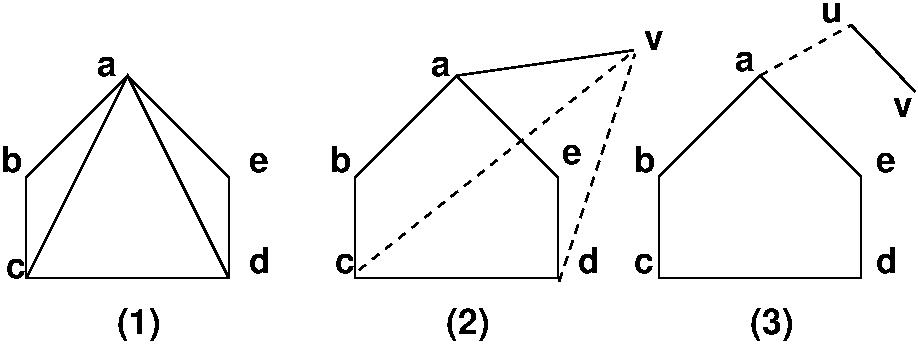}
\caption{Neighbourly set with a pentagon (Proof of Lemma 2 and
Corollary 1)}
\end{figure}

{\textbf{Corollary 1:}}
If a special neighbourly set $A^*$ of a quadrilateral-free graph $G$ has
a pentagon $abcde$, then there are no edges in $A^*$ other than the five
edges forming the pentagon.

{\textbf{Proof:}} We know from the proof of the lemma, that we can not
have an edge in $A^*$ 

(1) between two vertices of the pentagon

(2) between one vertex of the pentagon, and one other vertex outside the
pentagon

Thus the only possibility which remains is that the edge $p$ is between
two vertices, (say) $u$ and $v$, both of which are distinct from any of
the pentagon vertices (see Figure 2).

Then $p$ and (say) edge $ab$ are both adjacent to an edge (say) $xy$
(else they both cannot be together in the neighbourly set). But as $A^*$
is a special neighbourly set, the edge $xy\in A^*$. From the second case
in the proof of lemma, we saw that such an edge can not exist. Thus, the
corollary follows. []

{\textbf{Lemma 3:}} If a special neighbourly set $A^*$ is cycle-free,
then all paths of $A^*$ are of length at most three.

{\textbf{Proof:}} Let us assume that there is a path, $a-b-c-d-e$ of
length four in $A^*$. As edges $ab$ and $de$, do not share an end point,
and as they are in a special neighbourly set, there must be an edge (say)
$xy$ adjacent to both $ab$ and $de$.  The edge $xy$ along with (part of)
the path $a-b-c-d-e$ will form a cycle contradicting our assumption that
$A$ is cycle free. []

Thus special neighbourly sets, if cycle free, can never have diameter
more than three.

To prove, the corresponding result for general neighbourly sets, we use
following results (see [11] for proof):

{\textbf{Lemma 4:}} [[11]] If $A$ is a neighbourly set of a
quadrilateral-free graph $G=(V,E)$, then

(1) If edges $xy_1, xy_2, x'y_1', x'y_2'$ are in $A$, and if all $x, x',
y_1, y_2, y_1', y_2' $ are distinct, then $xx'\in E$ is an edge of $G$.

(2) If three or more edges of $A$ are incident at a vertex $v\in V$, then
all edges of $A$ will have at least one endpoint adjacent to $v$.

{\textbf{Lemma 5:}} Let $A$ be a neighbourly set of a quadrilateral-free
graph $G$. Then there cannot be a chordless path of length more than four
in $A$ (actually in the subgraph $(V[A],A)$ where $V[A]$ denotes the
endpoints of edges of $A$).

{\textbf{Proof:}} Suppose there is a path $abcdef$ of length five.  Then
$a,b,c,d,e,f$ are distinct vertices and $ba, bc, ed,ef$ are edges in the
neighbourly set $A$. So by Lemma 4, $b$ and $e$ must be adjacent. This
will lead to a quadrilateral $bcde$.  []

{\textbf{Lemma 6:}} Let $A$ be a neighbourly set of a quadrilateral-free
graph $G$.  If there is a path $abcde$ of length four in $A$ exactly one
of the following edges must be in $G$: \\ \textbf{Type 1:} Edge $ae$ \\
\textbf{Type 2:} Edge $bd$

{\textbf{Proof:}} As $ab$ and $de$, do not share an endpoint, there must
be an edge connecting them. 

As presence of either edge $ad$ or $be$ will result in a quadrilateral,
the only possibility which remains is of edges $ae$ and $bd$.  As
presence of both these edges will result in a quadrilateral, the lemma
follows. []

{\textbf{Corollary 2:}} If $A$ is an neighbourly set of a
quadrilateral-free graph $G$, then there are no cycles of length more
than five in $A$.

{\textbf{Proof:}} A cycle of length more than five will have a path of
length at least five which, we know from Lemma 5, is not possible. []

\section{Special Neighbourly Sets} 

Our algorithm for special neighbourly sets is based on the following
theorem.

{\textbf{Theorem 1:}} Let $A^*$ be a special neighbourly set in a
quadrilateral-free graph $G$. Let $E_u$ denote the set of edges of $G$
which are incident at $u$. Then at least one of the following
propositions holds:

(1) $A^*$ has an edge $ab$ such that $A^* \subseteq E_a \cup E_b $
 
(2) $A^*$ has a triangle $\triangle abc$ such that $A^* \subseteq E_a
\cup E_b \cup E_c$
 
(3) $A$ is a pentagon or $C_5$.

{\textbf{Proof:}} Let $A^*$ be a special neighbourly set of a
quadrilateral-free graph $G$. Then if $A^*$ is cycle-free then by Lemma
3, all paths are of length at most three. Thus in special neighbourly
sets, there will be an edge (say) $ab$ such that all edges of $A^*$ are
adjacent to it (see Figure 3). Thus $A \subseteq E_a \cup
E_b$.

\begin{figure}[h!] 
\centering
\includegraphics[width=1.9in]{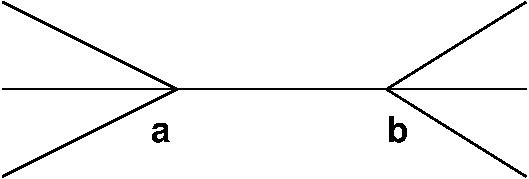}
\caption{Neighbourly set without cycles}
\end{figure}

Next consider the case when $A^*$ has a cycle. We know from Corollary 2,
that the largest cycle can be of length five. If there is a cycle of
length five, then by Corollary 1, $A^*$ has no other edges. 

As the graph $G$ is quadrilateral-free, there can be no cycles of length
four. The only possibility that remains is that one of the cycles is a
triangle (say $\triangle abc$). In this case, by Lemma 1, $A^* \subseteq
E_a \cup E_b \cup E_c$.

Thus the theorem follows. []

In our proposed algorithm, we find a special neighbourly set of largest
cardinality satisfying condition (1) and another satisfying condition (2)
of Theorem 1, and pick the one having larger cardinality. If its
cardinality is less than five, we look for a neighbourly set satisfying
condition (3).  Less informally, the algorithm is

(1) We first find the degree $d_v$ of each node $v$ in $O(m+n)$ time
(just count the number of edges incident at each node). Note that $d_a$
is $|E_a|$.

(2) To find a special neighbourly sets satisfying (1), we examine each
edge $ab$, and find $|E_a \cup E_b| = |E_a| + |E_b| -1 $.

We pick the edge $uv$ for which $|E_u|+|E_v|$ is
maximum. The cardinality of special neighbourly set $A_1^*$ will be 
$|E_u|+|E_v|-1$. This step clearly takes $O(m)$
time.

(3) We enumerate all triangles (using the algorithm of Section 2), and
choose the triangle $uvw$ for which
$|E_u|+|E_v|+|E_w|$ is maximum. The
cardinality of the special neighbourly set $A_2^*$ will be
$|E_u|+|E_v|+|E_w|-3$.

(4) If cardinality of neighbourly sets found in first two cases are both
$4$ or less, then we search for a pentagon. This step is described in
more detail later.

Step 1 and Step 2 can be implemented in $O(n+m)$ time. We saw in Section
2 that Step 3 can be implemented in $O(n^2)$ time. We next show that Step
4 can also be implemented in linear time.

We basically check for each vertex $v$, in turn, if it is a part of a
pentagon by taking all possible paths of length $5$ starting with vertex
$v$ and testing if it is a pentagon.  

Recall that we are looking for pentagons only when the cardinality of
$A_1^*$ (and $A_2^*$) is less than five. This means that maximum degree
of the graph is also less than five.  Therefore at each vertex, we will
have at most four different edges to choose from and since we limit the
length of the path to five, we take at most a constant number of steps
(less than $4^5$) to check for each vertex if it is part of a pentagon.
Thus, time spent at each vertex is $O(1)$.  Therefore, in $O(n)$ time, we
can find if there are any pentagons in the graph. 

Thus, our proposed algorithm to find a maximum cardinality special
neighbourly set in a quadrilateral-free graph takes $O(n^2)$ time
algorithm. In case of planar quadrilateral-free graphs, the algorithm
takes linear time.

The bottleneck is the algorithm to enumerate all triangles in a
quadrilateral-free graph.  An $O(m+n)$ time algorithm for finding all
triangles in quadrilateral-free graphs will also result in a linear time
solution for the maximum special neighbourly set problem in
quadrilateral-free graphs.

\section{Properties of General Neighbourly Sets} 

We next discuss some properties of general neighbourly sets in 
quadrilateral-free graphs. The first three lemmas directly follow from
the first part of Lemma 4. To avoid cumbersome notation, we will use $A$
both for the subset of edges and also for the edge induced subgraph
$(V[A],A)$ (here $V[A]$ is the set of endpoints of $A$). 

{\textbf{Lemma 7:}} Let $A$ be a neighbourly set of a quadrilateral-free
graph $G$. If $A$ has a path $L$ of length four then all paths having no
vertex in common with $L$ are of length at most one.

{\textbf{Proof:}} If $abcde$ is a path of length four, then edges
$ab,bc,cd,de$ will be in $A$. If $fgh$ is a path of length two in $A$.
Then edges $fg,gh$ will be in $A$. Applying (first part of) Lemma 4, for
edges $ab,bc,fg,gh$, we can infer presence of edge $bg$ in $G$. Again
applying Lemma 4, for edges $cd,de,fg,gh$, we see that edge $gd$ is
present in $G$. But, these edges result in a quadrilateral $bgdc$, a
contradiction. []

{\textbf{Lemma 8:}} If there is a path $L$ of length three in a
neighbourly set $A$ of a quadrilateral-free graph $G$, then there can be
at most one more path, having no vertex in common with $L$, of length
more than one in $A$.  If such a path exists it has length exactly two.

\begin{figure}[h!] 
\centering
\includegraphics[width=3.9in]{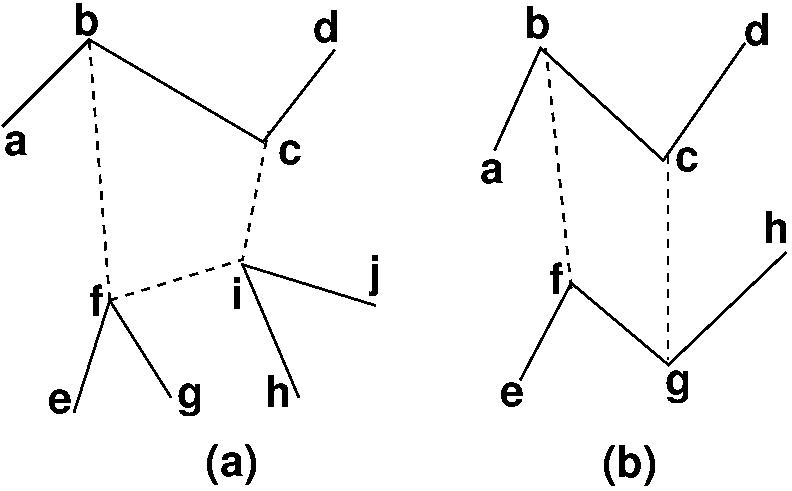}
\caption{Neighbourly set containing a path of length three}
\end{figure}

{\textbf{Proof:}} Let $abcd$ be a path of length three in $A$. Let us
assume that there are two other vertex disjoint paths of length more than
one. Let $efg$ and $hij$ be (part of) these paths (see Fig 4(a)). 

By first part of Lemma 4, as edges $ab,bc,ef,fg$ are in $A$, edge $bf$
must exists in the graph $G$.  Again from presence of edges
$bc,cd,hi,ij$, we can infer the presence of edge $ci$ in $G$.  Finally,
from presence of edges $hi,ij,ef,fg$ we can infer the presence of edge
$fi$ in $G$. Thus, we get a quadrilateral $bcif$, a contradiction. Hence,
we can have at most one more vertex disjoint path of length more than
one.

Presence of another vertex disjoint (part of) path of length $3$, say
$efgh$, by a similar argument will result in a quadrilateral $bcgf$ (see
Figure 4 (b)). Thus, the lemma follows. []

{\textbf{Lemma 9:}} Let $G$ be a quadrilateral-free graph, and $A$ a
neighbourly set of $G$.  Then there cannot be more than three vertex
disjoint paths of length two in $A$.
 
{\textbf{Proof:}} Let us assume that we have four vertex disjoint paths
of length three (say) $awA,bxB,cyC,dzD$.

\begin{figure}[h!] 
\centering
\includegraphics[width=2.9in]{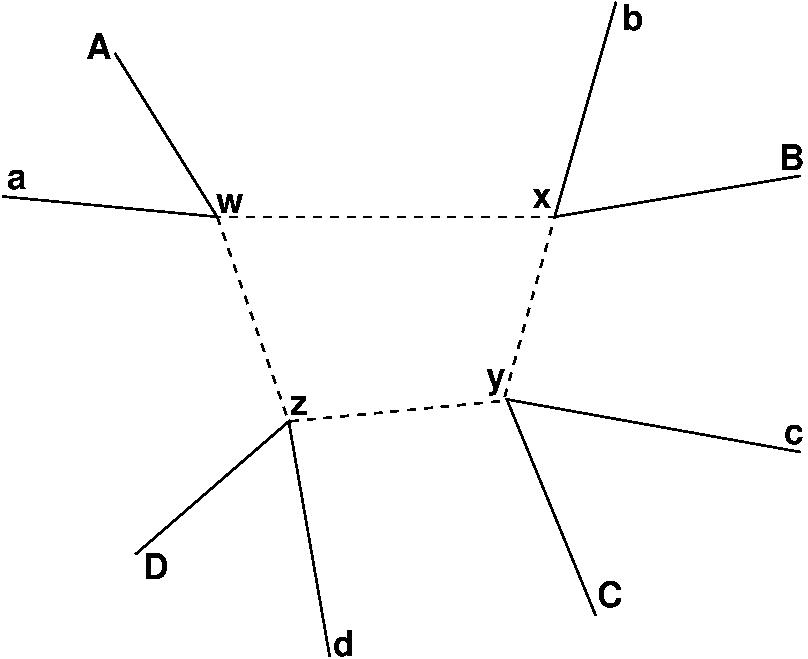}
\caption{Four paths of length two in an neighbourly set}
\end{figure}

Then, by an argument similar to that in Lemmas 8 and 9, we can infer
quadrilateral $wxyz$. []

{\textbf{Lemma 10:}} Let $A$ be a neighbourly set of a quadrilateral-free
graph $G$.  If $A$ has a vertex $x$ along with multiple paths (spokes) of
length one or two emanating from $x$, then $A$ contains at most three
such spokes of length two.

{\textbf{Proof:}} Let us assume that $A$ has four ``spokes'' of length
two (See Figure 6):\\
$xaa',xbb',xcc',xdd'$

As each of $a'axbb'$, $a'axdd'$, $b'bxcc'$ and $c'cxdd'$ form a path of
length four, so by Lemma 6, there are edges $a'b', a'd', b'c', c'd'$ of
Type 1 or edges $ab,ad,bc,cd$ of Type 2 in respective paths.

If all these edges are of type 1, then we will get a quadrilateral
$a'b'c'd'$. Thus, at least one of these has to be an edge of type 2.
Without loss of generality, let us assume this is edge $ab$. 

Observe that $xab$ forms a triangle. 

\begin{figure}[h!] 
\centering
\includegraphics[width=2.6in]{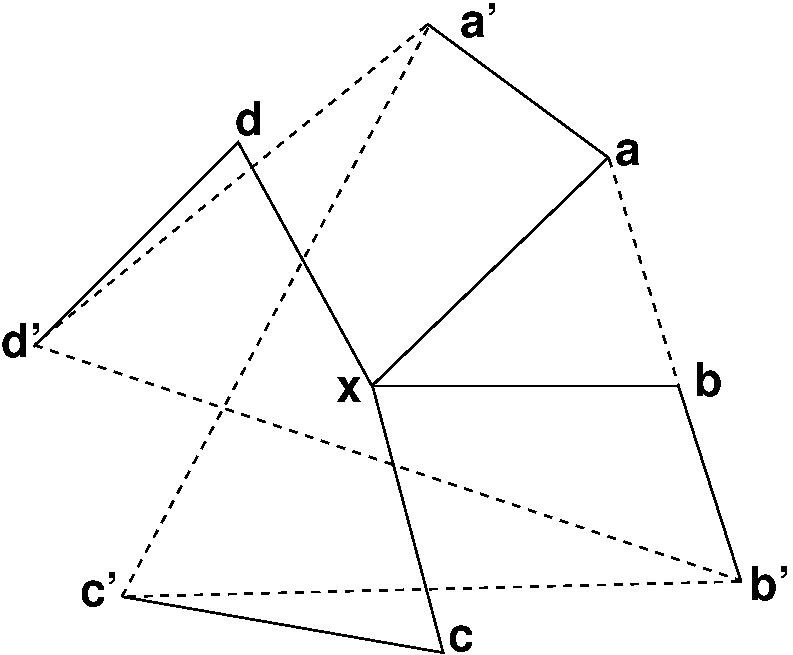}
\caption{Neighbourly set with four spokes}
\end{figure}

Let us consider following paths of length $4$ , one by one:\\
$a'axcc',  a'axdd',  b'bxcc', b'bxdd'$

{\textbf{$a'axcc'$:}} By Lemma 6, either edge $a'c'$ or edge $ac$ must be
present. But as presence of edge $ac$ results in quadrilateral $cabx$ (a
contradiction), the edge $a'c'$ must be present.

{\textbf{$a'axdd'$:}} Again, by Lemma 6, we must have either edge $a'd'$
or edge $ad$. But as presence of edge $ad$ results in quadrilateral
$dabx$, edge $a'd'$ must be present.

{\textbf{$b'bxcc'$:}} As presence of edge $bc$ results in quadrilateral
$cbax$, edge $b'c'$ must be present.

{\textbf{$bb'xdd'$:}} As edge $bd$ results in quadrilateral $dbax$, edge
$b'd'$ must be present.

Now as edges $a'c',a'd',b'c',b'd'$ result in quadrilateral $a'c'b'd'$, we
get a contradiction. []

For the case, when $A$ contains a cycle of length five, or a pentagon, we
have the following result:

{\textbf{Lemma 11:}} If $A$ contains a pentagon (say) $abcde$, then $A$
can have at most one other edge (say) $a'b'$ with $a',b'$ different from
pentagonal vertices.
 
{\textbf{Proof:}} If $A$ contains a pentagon (say) $abcde$, then from
Lemma 2, no other edge of $A$ can be incident at any of the pentagonal
vertices $\{a,b,c,d,e\}$. However, $A$ can have an edge (say) $a'b'$ with
$a',b'$ distinct from pentagonal vertices (see Figure 7).

\begin{figure}[h!] 
\centering
\includegraphics[width=4.4in]{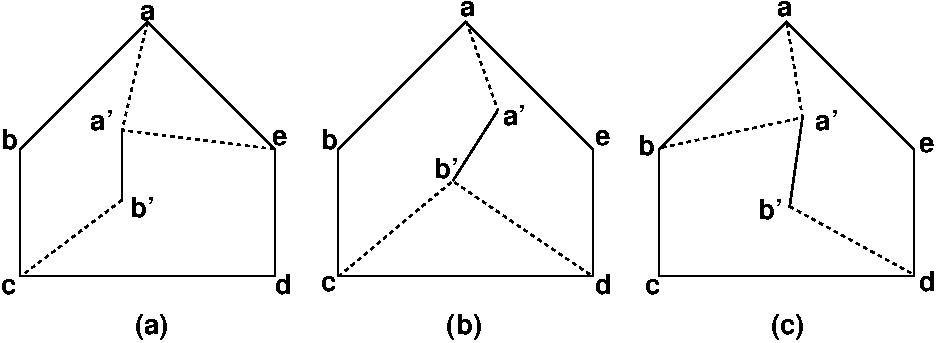}
\caption{Neighbourly set with a pentagon and an edge not touching the
pentagonal vertices} 
\end{figure}

As $a'b'$ and $ab$ are both present in the neighbourly set $A$, they must
be connected by an edge of $E$ (which by Lemma 2, is not in $A$). Without
loss of generality let $aa'$ be this edge. Observe that we can neither
have edge $b'e$ (we will get quadrilateral $aa'b'e$) nor edge $b'b$ (as
we will get quadrilateral $aa'b'b$).  Similarly edges $ca'$ and $da'$ are
prohibited (we get quadrilateral $aa'cb$ or $aa'de$). 

As $ed$ and $a'b'$ are both in $A$, we must either have the edge $a'e$ or
the edge $db'$ to join them. Similarly either edge $ba'$ or $cb'$ must be
present to join $bc$ and $a'b'$.

Let us first consider the case when edge $a'e$ is present in $G$ (see
Figure 7(a)). Edge $db'$ is not present as it will lead to quadrilateral
$a'b'de$.  As presence of edge $ba'$ will lead to quadrilateral $ba'ea$,
edge $cb'$ must be present to connect $bc$ and $a'b'$. 

If edge $a'e$ is not present in $G$, then we need edge $db'$ to join $ed$
and $a'b'$. To joint $bc$ and $a'b'$ we need either the edge $cb'$ (see
Figure 7(b)) or the edge $ba'$ (see Figure 7(c)).

Note that both edges $cb'$ and $ba'$ can not be present, as this will
lead to quadrilateral $ba'b'c$. 

All the three possibilities listed above (graphs (a), (b) and (c) in
Figure 7) are actually isomorphic to each other (one side of pentagon
forms a triangle with either $a'$ or $b'$).  Let us call this
configuration a {\textbf{pentagonal pseudo-prism}}.

If $A$ has a path of length more than one, which does not touch any of
the pentagonal vertices, then each edge in the path has to form a
pentagonal pseudo-prism arrangement with the pentagon. It can be observed
that this is not possible in a quadrilateral-free setup.

Finally, assume there are two such disjoint edges, say $a'b'$ and
$a''b''$.  Without loss of generality, we can assume that $a'b'$ is
present with the pentagon in the middle configuration of Figure 7.

Vertex $b''$ can form a triangle in $G$ with either side $cd$, or $bc$ or
$ab$ (see Figure 8). The other two cases of sides $ae$ and
$cd$ are symmetric (mirror images). 

\begin{figure}[h!] 
\centering
\includegraphics[width=4.7in]{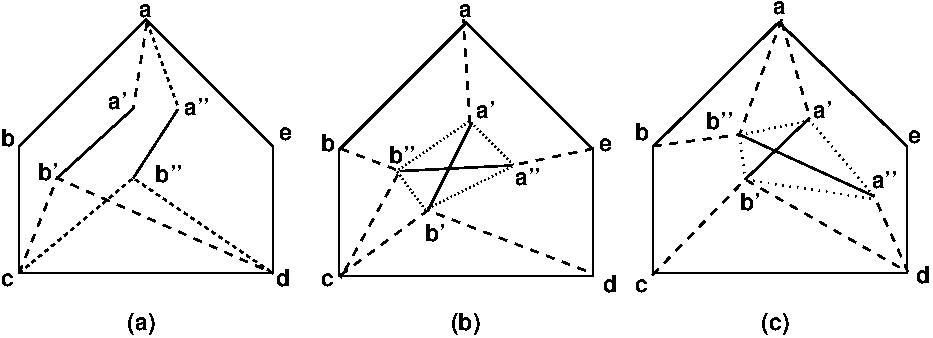}
\caption{Neighbourly set with pentagon and two other edges}
\end{figure}

If triangle $\triangle b''cd$ is present (see Figure 8(a)), then vertices
$cb'db''$ will forms a quadrilateral.

As $a'b'$ and $a''b''$ are both in $A$, there must be an edge connecting
them. This edge has to be from $a'a'',a'b'',b'a'',b'b''$.

Next consider the case, when triangle $\triangle b''bc$ is present. As
edge $a'b''$ will leads to quadrilateral $aa'b''b$, edge $b''b'$ to
$bb''b'c$, edge $b'a''$ to $b'a''ed$ and finally edge $a'a''$ to
quadrilateral $a'a''ea$ (see Figure 8(b)), this configuration is also not
possible.

Finally consider the case when we have triangle $\triangle abb''$.  In
this case too, edge $a'a''$ will result in quadrilateral $b'a'a''d$, edge
$a''b'$ in $a''dcb'$, edge $b'b''$ in $b'cbb''$ and finally edge $a'b''$
in quadrilateral $a'b''ba$.  (see Figure 8(c)), the lemma follows. []

\section{Characterising Neighbourly set} 

We will see that in the algorithm for general neighbourly sets, there is 
only one additional configuration, the configuration discussed in Lemma
10, which was not present in special neighbourly sets. 

We basically, show in the next section (Section 7) that all other
neighbourly sets can be ignored while looking for the maximal cardinality
neighbourly set as the graph also contains another neighbourly set of at
least the same cardinality, which is not ignored by our proposed
algorithm. 

Let $A$ be any neighbourly set in a quadrilateral-free graph $G$. Let us
first consider the case when $A$ has a cycle. We know from Lemma 2, that
this cycle will be either of length three or of length five. Further, we
know from Lemma 1 that if $A$ contains a cycle of length three or a
triangle (say) $\triangle abc$ then, $A \subseteq E_a \cup E_b \cup E_c$.

If $A$ contains a pentagon (say) $abcde$, then from Lemma 2, no other
edge of $A$ can be incident at any of the pentagonal vertices
$\{a,b,c,d,e\}$. Thus, either $A$ is just this pentagon or $A$ may have
an edge $a'b'$ with $a',b'$ distinct from pentagonal vertices as in Lemma
11. But in the later case, as we saw in proof of Lemma 11, there will be
triangle (e.g. $\triangle a'ae$ in Figure 7 (a)). This triangle with
edges adjacent to it will also give another neighbourly set with six
edges, the same as in the neighbourly set $A$, hence there is no
advantage in considering this set.

Thus, to summarise if a maximum cardinality neighbourly set $A$ has at
least one cycle, then we may assume that one of the following conditions
hold:

(1) There is a triangle $abc$ such that $A\subseteq E_a \cup E_b\cup E_c$
(necessarily including all the triangle edges)

(2) There is a pentagon such that $A$ is a set containing all its edges

Hence, in this case, we can proceed as in the algorithm for special
neighbourly set. Next let us consider the case when the neighbourly set
$A$ is cycle free.

First consider the case, when $A$ has two or more vertices of degree at
least three. In this case, from Mahdian (case 1 in proof of Theorem 4 of
[11]) it follows that as $A$ is cycle free (and hence triangle free) it
is of the form $E_u \cup E_v$ for some edge $uv$. 

Next consider the case when there is exactly one vertex (say $x$) of
degree three or more in $A$. From Lemma 4, every edge in $A$ has at least
one endpoint adjacent to $x$. Thus, the resultant neighbourly set will
contain spokes of length one or two with $x$ as centre (see Figure
9).

As all vertices, except $x$ are of degree $1$ or $2$ in (i.e., their 
degree in the subgraph induced by) $A$, no vertex $v$ adjacent to $x$ can
be be part of more than one spoke of length two.  Excluding edge $xv$ in
our neighbourly set $A$ will allow at most one other edge incident at $v$
to be included in neighbourly set.  But as this will give no advantage---
the neighbourly set with edge $xv$ will at least be of same cardinality,
hence we need not consider the possibility of excluding edges like $xv$ 
any further.

\begin{figure}[h!] 
\centering
\includegraphics[width=1.9in]{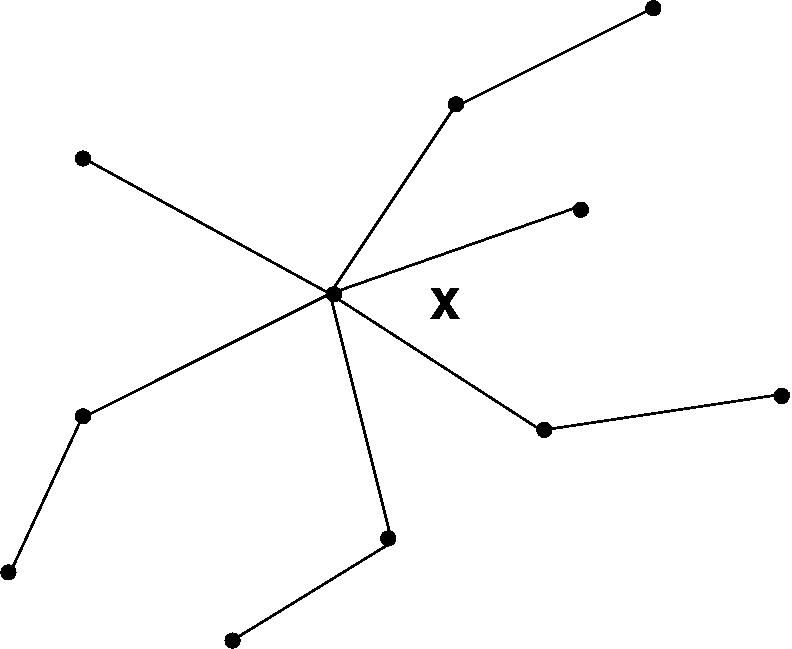}
\caption{Neighbourly set with exactly one vertex of degree more than two}
\end{figure}

From Lemma 10, there are at most three spokes of length two in $A$. As
even with three spokes of length two, neighbourly set will have a
cardinality at most $|E_x|+3$, we need to look at this case, only when
degrees of all neighbours of $x$ in graph $G$ is three or less (if a
neighbour $y$, is of degree four, then neighbourly set $E_x \cup E_y$
will be of size $|E_x|+3$).

Neighbourly set with just one spoke $xuv$ of length two will be an
neighbourly set of the form $E_x \cup E_u$, which we have already
considered. Thus, we need to look at cases, when the number of length two
spokes is either two or three.

Let $xaa'$ and $xbb'$ be (any) two spokes of length two. As in the proof
of Lemma 10, these spokes together form a path $b'bxaa'$ of length four,
and thus we must either have a connecting edge $b'a'$ (of Type 1) or an
edge $ba$ of Type 2 (see Lemma 6).

In case the edge is of Type 2 (for example $ba$), the the neighbourly set
with the triangle $\triangle xab$ will be of cardinality $|E_x \cup E_a
\cup E_b| \geq |E_x| +3$.  Thus, we need to consider only the case when
all the edges are of Type 1. Moreover, if there are three length $2$
spokes of Type 1, then there will be a triangle in $G$ between the ends
of the three spokes in the neighbourly set.

To summarise, neighbourly sets which we need to look at are only of
following types:

(1) $E_a \cup E_b\cup E_c$ for some a triangle $\triangle abc$.

(2) A pentagon.

(3) $E_u \cup E_v$ for some edge $uv$.

(4) There is a vertex with two or three spokes of length two and any
number of spokes of length one. When the number of spokes is three, we
need to consider only the case where the connecting edge between any two
different spokes are of type 1 and in this case, there is a triangle
between the ends of the three spokes.

For each type of neighbourly set as listed above, we find the instance of
maximum cardinality, and then compare all their cardinalities to find the
one with overall maximum cardinality. This will be the required maximum
cardinality neighbourly set. The algorithm of Section 4 for maximum
cardinality special neighbourly set, discussed all but the last case.

The special neighbourly set found will be the required set, unless there
is a neighbourly set of larger cardinality in Case 4. In particular, we
need to look at the last case, and search for three length two spokes
(respectively two length two spokes) only when the cardinality of the
special neighbourly set is less than $|E_\alpha|+3$ (respectively
$|E_\alpha|+2$), where $\alpha$ is the vertex of maximum degree in $G$.

Let us assume that we want to search for three (respectively two) length
two spokes at vertex $x$. We need to search, only if all neighbours of
$x$ have degree at most three (respectively two) in $G$ otherwise $|E_x
\cup E_y|> |E_x|+3$ (respectively $|E_x \cup E_y|> |E_x|+2$).

In the case of two length two spokes, we need to identify two spokes
(say) $a'ax$ and $b'bx$ such that (Type 1) edge $a'b'$ is in $G$.  In
case of three length two spokes, we need to identify three spokes (say)
$a'ax$, $b'bx$ and $c'cx$ such that all (Type 1 edges) edges of triangle
$\triangle a'b'c'$ are in $G$.

Let $N[x]=\{u_i\}_{i=1}^{d_x}$ be the vertices adjacent to $x$ in the
graph $G$ (here $d_x$ is the degree of $x$ in $G$). Let\\
$N^2[x]=\{v_i| (u,v_i)\in E \mbox{ for some } u\in N[x]\}$
be the vertices adjacent to vertices of $N[x]$.

We have to search for two spokes, only when all vertices in $N[x]$ are of
degree at most two; which means each $u_i$ is adjacent to at most one
vertex other than $x$. Thus, there will be at most $d_x$ (degree of $x$)
vertices in $N^2[x]$. Our search will be over, if there is an edge
between any pair of vertices in $N^2[x]$. We can easily check this by
looking at all $d_{x}^2$ entries in adjacency matrix of $G$.  If we find
an edge, we get a neighbourly set of size $|E_x|+2$. 

In case of three spokes, all vertices in $N[x]$ are of degree at most
three; which means each $u_i$ is adjacent to at most two vertices other
than $x$. Thus, there will be at most $2d_x$ (degree of $x$) vertices in
$N^2[x]$. Our search will be over, if there is a triangle (say)
$\triangle v_jv_kv_l$ in $N^2[x]$, such that each $v_i$ is adjacent to a
different $u_i$. To check this, we construct a new adjacency matrix for
the subgraph induced by $O({d_x})$ vertices of $N^2[x]$ We do this by
extracting $O(d^2_x)$ entries from the original adjacency matrix of $G$
in $O({d_x}^2)$ time. 

As this is a subgraph of a quadrilateral-free graph, it is also
quadrilateral free, hence using the algorithm of Section 2, we can
enumerate all triangles in $O({d_x}^2)$ time. For each triangle,
$\triangle v_jv_kv_l$, we check, in $O(1)$ time, if each $v_i$ is
adjacent to a distinct $u_i$.  If we find such a triangle, we have a
neighbourly set of size $|E_x|+3$, consisting of edges in $E_x$ along
with the edges of the three spokes whose end points formed the triangle.

Thus, we can check whether for spokes at vertex $x$, in both cases, in
$O(d^2_x)$ time. If we carry out above procedure, in decreasing order of
degrees, the total time taken will be at most $O(\sum\limits_{i=1}^n
d_i^2)$, which is $O(n^2)$ time for quadrilateral-free graphs [10]. Thus,
we can find a maximum cardinality neighbourly set in $O(n^2)$ time.

So the algorithm runs in overall $O(n^2)$ time, as compared to the $O(n
m^{10})$ time taken by the algorithm given in [11].

\section{Degree of each vertex in $A$ is at most $2$} 

If each vertex is of degrees one or two in $A$, then as $A$ is
cycle-free, it will just be a collection of paths. We will show, in each
such case, $G$ will always contain another neighbourly set, of an earlier
type, having at least the same cardinality. Thus, neighbourly set of this
form can be safely ignored. By Lemma 5, we know that
the length of any path in $A$ is at most four. 

\subsection{Length four Path}

Let us first consider the case when the longest path in $A$ is of length
four. We know from Lemma 7, that all other paths in
$A$ are of length at most one (disjoint edges).  Let the length four path
be $abcde$. We know from Lemma 6, that $G$ contains either
edge $bd$ or edge $ae$.

Presence of edge $bd$ in $G$ will result in triangle $\triangle bcd$ in
graph $G$. Any other edge (say $fg$) in 
the neighbourly set $A$ will have to be joined to edge $bc$ by an edge
$fb$ or $gb$ (incident at $b$) or by edge $fc$ or $gc$ (incident at $c$).
Thus all such edges will be in the set $E_b \cup E_c \cup E_d$. Thus,
$E_b \cup E_c \cup E_d$ will always be a neighbourly set of larger
cardinality than $A$. 

Hence, we need to only consider the case when edge $bd$ is not present.
Presence of edge $ae$ will result in pentagon $abcde$.  Simply taking
this pentagon, will result in a neighbourly set of size five. Thus, we
need consider the case further when there are at least two other disjoint
edges in the neighbourly set.

Let $fg$ be any such edge.  As edge $fg$ must be joined to edges $bc$ and
$cd$, either one of the end point (say $f$) is joined to $c$ (i.e., edge
$fc$ is in $G$), or the edge $fg$ must be joined to both $b$ and $g$. As
presence of both edges $fb$ and $fd$ will result in quadrilateral $fbcd$,
we must have the pair of edges $fb$ and $gd$.

Let us first consider the case (Case 1) when edge $fc$ is in $G$. In this
case edges $ab$ and $de$ have to be joined to edge $fg$.  We can neither
have edge $bg$ nor edge $dg$ as their presence will lead to quadrilateral
$bcfg$ or $cdgf$.  Similarly, we can neither have edges $af$ nor $ef$ as
their presence will lead to quadrilaterals $abcf$ or to $cdef$. 

Thus, we are left with edges $ag, bf, eg, df$.

Presence of either edge $bf$ or $df$ (see Figures 10 (a) and
(b)), will result in either triangle $\triangle bcf$ or in $\triangle
dcf$. Observe that $E_b\cup E_c \cup E_f$ or $E_d\cup E_c \cup E_f$ will
be a neighbourly set of size six. Let us call this as Case 1(a). If any
other edge (say) $hi$ is present in $A$, then it has to be joined to edge
$bc$ (or edge $cd$ in the second case) by an edge incident at either $b$
or $c$ ($c$ or $d$ in the second case). Thus, all other edges of $A$ will
also be in the set $E_b\cup E_c \cup E_f$ or $E_d\cup E_c \cup E_f$
(which we have dealt with earlier). Thus, we can ignore this case.

\begin{figure}[h!] 
\centering
\includegraphics[width=3.9in]{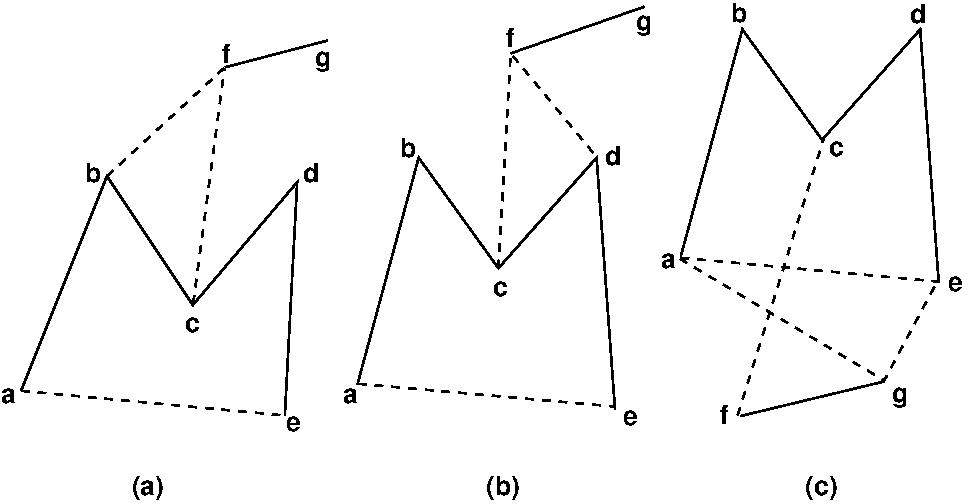}
\caption{$c$ is adjacent to $f$}
\end{figure} 

If edges $bf$ and $df$ are both absent, then we must have the pair of
edges $ag$ and $eg$ to join $fg$ to both $ab$ and $ed$. However, presence
of both these edges will result in a triangle $\triangle aeg$.
Neighbourly set $E_a \cup E_e \cup E_g$ is of cardinality six (see Figure
10 (c)). We will call this as Case 1(b).

Observe that if we have another edge (say) $hi$, in Case 1(b), then
presence of both triangles $\triangle aeg$ and $\triangle aig$ will
together result in quadrilateral $agei$ (see Figure 11). Hence, there can
be at most one edge in Case 1(b). Unless there are edges in the remaining
case, Case 2, which we consider next, this case can also be ignored.

\begin{figure}[h!] 
\centering
\includegraphics[width=2.6in]{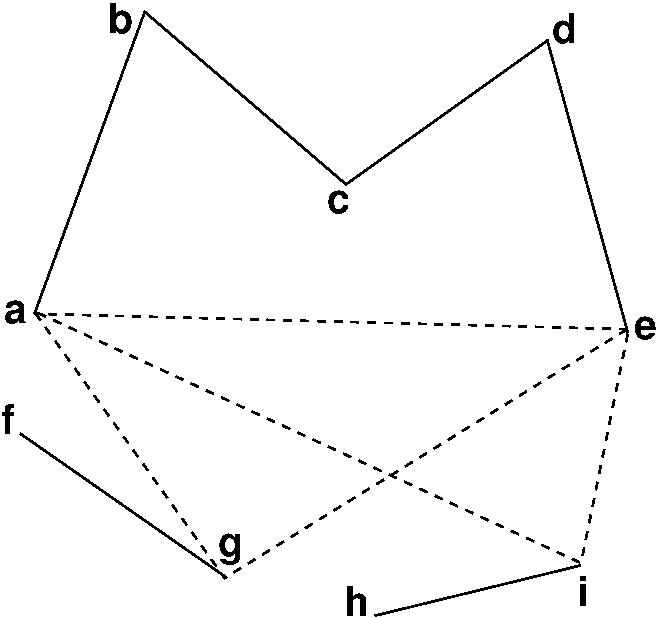}
\caption{More than one edge in configuration 1(a)} 
\end{figure}

Finally consider the case when neither edge $fc$ nor $gc$ is present.
Then $bc$ and $cd$ have to be connected to $fg$ by edges $bf$ and $dg$;
they cannot both be adjacent to the same point as that will result in a
quadrilateral.  If there are two edges say $f_1g_1$ and $f_2g_2$ in this
case, then we know we will have edges $f_1b, f_2b, g_1d, g_2d$.  Edge
$f_1g_1$ has to be connected to $f_2g_2$ either by $f_1f_2$ or by
$g_1g_2$ (for example, the edge $f_1g_2$ will result in a quadrilateral
$bf_1g_2f_2$).  Edge $f_1g_1$ or $f_2g_2$ will result in triangle
$\triangle bf_1f_2$ or $\triangle dg_1g_2$, resulting in a neighbourly
set of size six.  We next show that there can be at most two edges in
this case. 

\begin{figure}[h!] 
\centering
\includegraphics[width=3.2in]{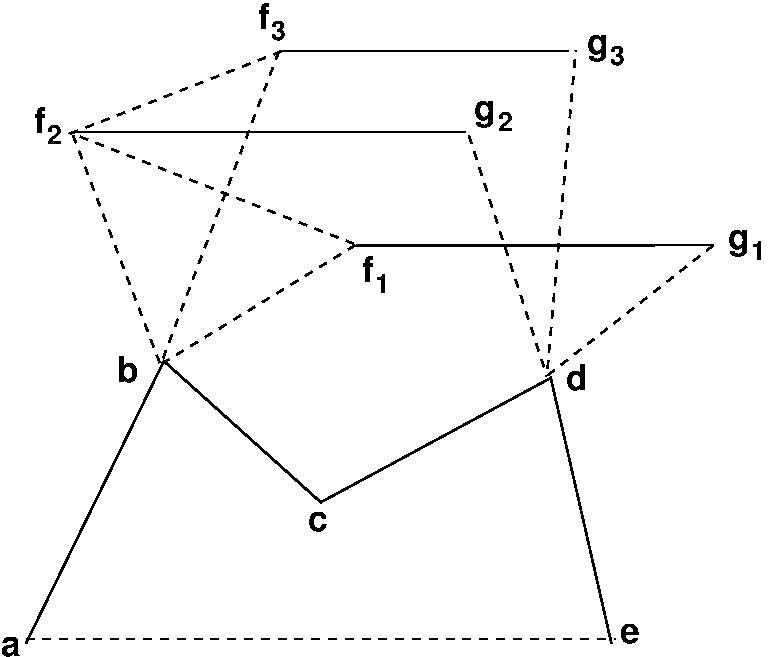}
\caption{More than two edges in configuration 2}
\end{figure}

Suppose there are three edges, say $f_1g_1$, $f_2g_2$ and $f_3g_3$.  Then
we will also have the corresponding edges $f_ib$ and $g_id$.  Observe
that $f_ig_i$ can not be connected to $f_jg_j$, (for $i\neq j$) by edge 
$f_ig_j$ as that will result in a quadrilateral $bf_ig_jf_j$.  Thus the
connecting edges are either $f_if_j$ or $g_ig_j$. Without loss of
generality, assume that $f_1g_1$ and $f_2g_2$ are joined by $f_1f_2$
(else replace ``$f$'' by ``$g$''). Now $f_1g_1$ can not be joined to
$f_3g_3$ by $f_1f_3$ as that will result in quadrilateral $bf_3f_1f_2$, 
hence they must be joined by the edge $g_1g_3$. If $f_2g_2$ is joined to
$f_3g_3$, by edge $f_2f_3$ or with edge $g_2g_3$ we will get
quadrilateral $bf_3f_2f_1$ or $dg_2g_3g_1$.  Thus, if we have only two
edges in Case 2 (and no edge in any other case), we can ignore this case.

The only possibility left is when there is one edge, say $hi$, in Case
1(b) and another edge, say $fg$ in Case 2. 

\begin{figure}[h!] 
\centering
\includegraphics[width=1.9in]{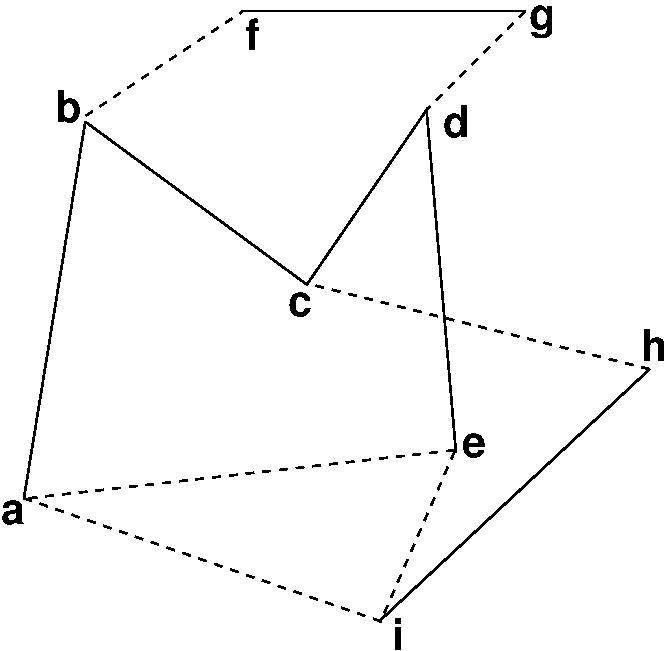}
\caption{No  edge in configuration 1(a)}
\end{figure}

As $hi$ is in Case 1(b), we have a triangle with one side as $ae$ and
third vertex as either $h$ or $i$. Without loss of generality, let us
assume that the triangle is $\triangle aei$. We also have edge $hc$. As
$fg$ is in Case 2, edges $fb$ and $gd$ are there.  Edge $hi$ has to be
joined to $fg$ (see Figure 12). 

As connecting edge $fi$ will result in quadrilateral $iabf$, edge $gi$ in
$iedg$, edge $fh$ in $hcbf$ and finally edge $gh$ in quadrilateral
$hcdg$, this case is also not possible.

\subsection{Length Three Paths}

Let us next consider the case when the longest path in $A$ is of length
three. From Lemma 8, there can be at most one more path of
length more than one; and if such a path exists, it is of length exactly
two.  

Let $abcd$ be a length three path in $A$. All edges of this path are in 
$E_b \cup E_c$.  As any length one path (edge) $ef$ in $A$ has to be
connected to $bc$, by an edge incident to either $b$ or $c$, thus for 
each such path in $A$, there is a corresponding edge in $E_b \cup E_c$.
Hence $E_b \cup E_c$ has cardinality at least four plus number of such
edges, the same as that of the neighbourly set $A$. 

\begin{figure}[h!] 
\centering
\includegraphics[width=1.6in]{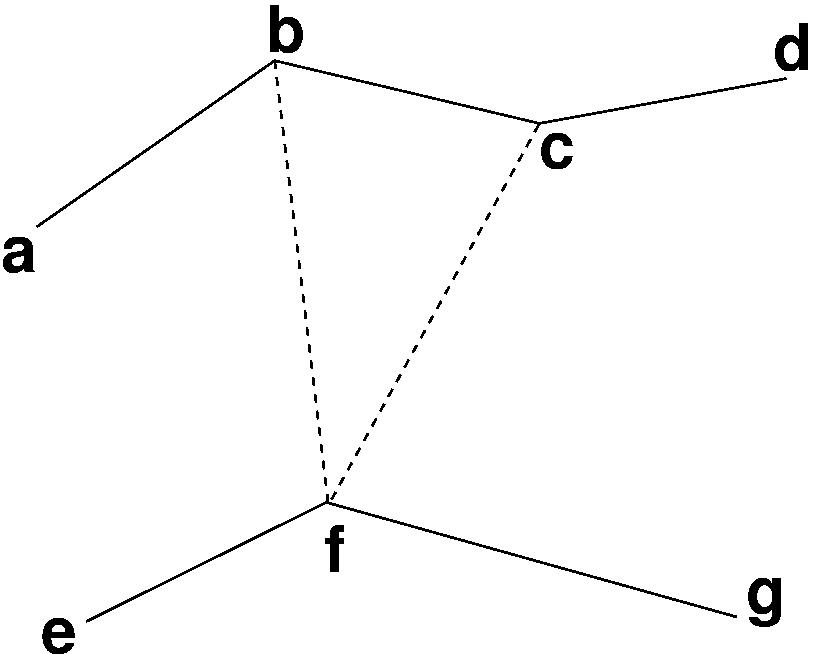}
\caption{Path of length 3 with a path of length two}
\end{figure}

If we have a path $efg$ of length two, then by applying Lemma 4 to
$ab,bc,ef,fg$ and to $bc,cd,ef,fg$, we can infer presence of edges $bf$
and $cf$ in $G$. Presence of these edges results in triangle $\triangle
bcf$. The neighbourly set $E_b \cup E_c \cup E_f$ is also of cardinality
seven (see Figure 14). Further, any edge $hi$ (path of length one) in $A$
has to be connected to $bc$, and corresponding to each such edge in $A$
there is an edge incident at either $b$ or $c$ in the graph $G$; this
edge will be in $E_b\cup E_c$ (see Figure 14). Hence, the set $E_b \cup
E_c \cup E_f$ is of higher cardinality.

\subsection{Length Two Paths}

Next consider the case when the longest path in $A$ is of length two. 
From Lemma 9 we know that there can be at most three paths of length two
in $A$.  

If there is only one length two path (say) $abc$, then every length one
path (edge) in $A$ has to be connected to edge $bc$. Thus, corresponding
to each edge in $A$, there is an edge incident at $b$ or $c$ in $E_b \cup
E_c$, connecting it to $bc$. Hence $E_b \cup E_c$ is another neighbourly
set of the same cardinality as that of $A$.  Hence, we can ignore this
case.

Before discussing other two cases (two and three length two paths),
observe that any path (say) $xy$ of length one, has to be connected to
both edges $ab$ and $bc$ of length two path $abc$. Without loss of
generality, let us assume that $x$ is connected to $ab$. Vertex $x$ can
be connected to $ab$ either by edge $xa$ or by edge $xb$ (see Figure
15).

If edge $xa$ is present in $G$, then as presence of edge $yb$ will lead
to quadrilateral $axyb$, edge $xy$ is joined to $bc$ by edge $yc$. Thus,
there will be a pentagon $abcxy$ in $G$.

\begin{figure}[h!] 
\centering
\includegraphics[width=2.3in]{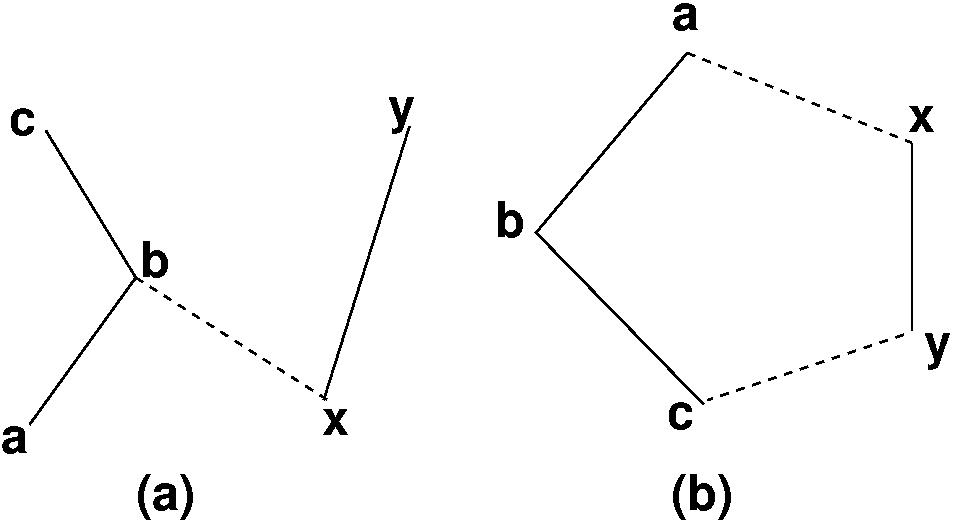}
\caption{One path length two and one path of length one}
\end{figure}

Next consider the case when there are two length two paths (say) $abc$
and $def$. Then applying Lemma 4, to edges
$ab,bc,de,ef$, we can infer presence of edge $be$ in $G$ (see
Figure 16). 

\begin{figure}[h!] 
\centering
\includegraphics[width=3.9in]{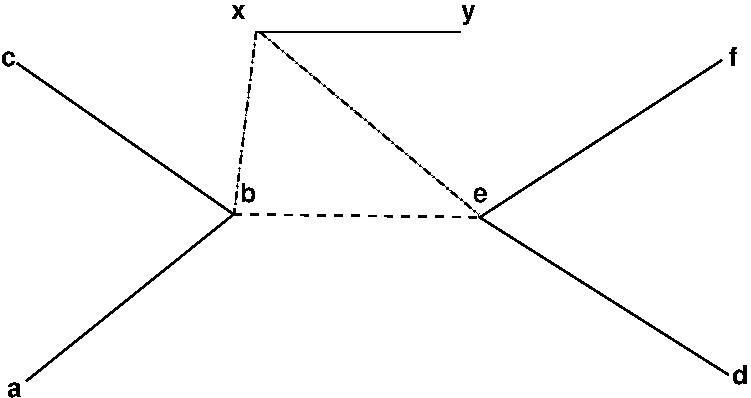}
\caption{Two paths of length two}
\end{figure}

Any length one path (say) $xy$ in $A$ has to be connected to edges
$ab,bc$ of path $abc$ either by using edge $xb$ or by using pair of edges
$xa$ and $yc$. Similarly edge $xy$ can be connected to edges $de,ef$ of
path $def$ by using either

(1) Pair of edges $xd$ and $yf$, or pair of edges $yd$ and $xf$, or

(2) Edge $xe$ or

(3) Edge $ye$.

Let us first consider the case when edge $xb$ is in the graph $G$. Then
as presence of edge $ye$ will result in quadrilateral $yebx$, of $xd$ in
quadrilateral $xdeb$ and of $xf$ in $xfeb$, the only possibility left is
of edge $xe$.

And if the pair of edges $xa$ and $yc$ is present, then the presence of
edge $xe$ or $ye$ will result in quadrilaterals $xeba$ or $ycbe$.  Thus,
in this case, either the pair $xd$ and $yf$, or the pair $yd$ and $xf$
will be present.  As both cases are similar (interchange $f$ and $d$),
let us assume that pair $xd$ and $yf$ is present.

Thus, to summarise, there are basically two possibilities:\\ $\bullet$
Case 1: Either edges $xb$ and $xe$ are present, or\\ $\bullet$ Case 2:
Edges $xa,yc,xd,yf$ are present. 

Observe that cardinality of $E_b \cup E_e \cup E_x$ is at least eight in
Case 1. And edges $xa,ab,bc,yc,yx$ will give a pentagon of cardinality
five in Case 2.

Let us assume that there is another edge (say) $x'y'$ in $A$. If both
edges are in the first case (Case 1) presence of edges $x'b,x'e$ and
$xb,xe$ will result in quadrilateral $exbx'$, hence this case is not
possible. Thus, there can be at most one edge (edge $xy$) in Case 1.

If both are in the second case, then due to $xy$, we will have edges
$xa,yc,xd,yf$. Again, without loss of generality we can assume that we
have edges $x'a,y'c$ (else interchange $x'$ and $y'$). Now as presence of
edge $x'd$ or $y'f$ will result in quadrilateral $x'dxa$ or $y'fyc$, we
must have edge $x'f$ and $y'd$.  As both $x'y'$ and $xy$ are in $A$, they
have to be joined by one of the following edges $xx',yy',xy',x'y$. Edges
$xx'$ or $yy'$ will result in quadrilaterals $xx'y'd$ or $yy'dx$.  And
edges $x'y$ or $xy'$ will result in quadrilaterals $x'yxa$ or $xy'cy$.
Thus, we can not have both edges in the second case. Hence, again there
can be at most one edge in Case 2 also.

\begin{figure}[h!] 
\centering
\includegraphics[width=2.6in]{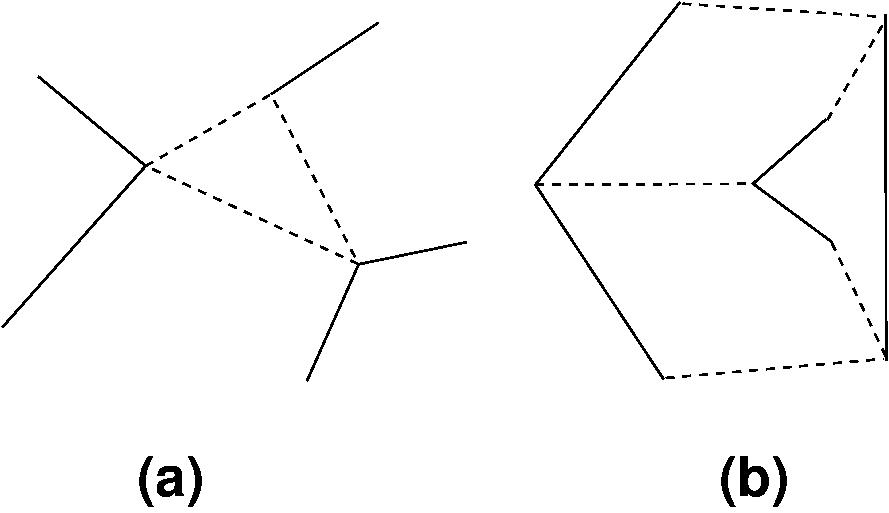}
\caption{Two paths of length two and one of length one}
\label{twoAndOne}
\end{figure}

If there are three edges (length one paths) in $A$, and two length two
paths, then at least two length one paths must be in the same case, which
is not possible. Hence, if there are two length two paths (say) $abc$ and
$def$ in $A$, we have following mutually exclusive possibilities:

$\bullet$ There is at most one length one path in $A$. In this case,
cardinality of $E_b \cup E_e$ will be at least five, which is at least
the same as cardinality of $A$.

$\bullet$ There are two length one paths in $A$. Then at least one path
(say $xy$) must be in Case 1, in this case cardinality of $A$ will be
six, which is less than eight, the cardinality of $E_b \cup E_e \cup
E_x$.

Hence, it is safe to ignore this case also.

Finally, let us consider the case when there are three length two paths
(say) $abc,def,ghi$. From proof of Lemma 9, the mid points $b,e,h$ of
these length two paths will be connected to each other, forming a
triangle $\triangle beh$ in the graph.  There will be at least nine edges
in $E_b \cup E_e \cup E_h$.

\begin{figure}[h!] 
\centering
\includegraphics[width=1.9in]{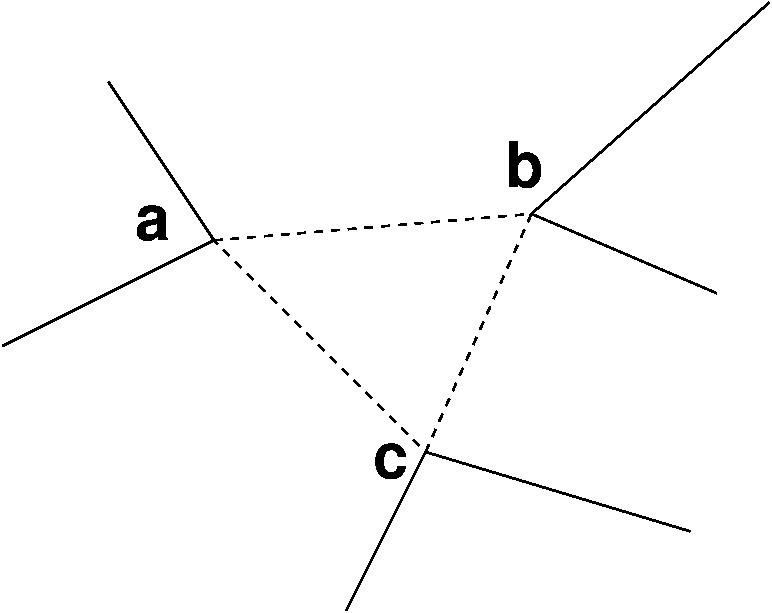}
\caption{Three paths of length two}
\end{figure}

We have just seen that even in presence of two length two paths, we can
have at most two paths of length one.  Cardinality of $A$ will be at most
$3\times 2+2=8$, less than nine edges in $E_b \cup E_e \cup E_h$. Hence,
this case can also be ignored.

\subsection{Length One Paths}

Let us finally consider the case when all paths in $A$ are of length one.
Let $ab$ be an edge in $A$.  Every other edge in $A$ has to be connected
to $ab$ at either $a$ or $b$. Thus, corresponding to every other edge
there is an edge in $E_a \cup E_b$ incident at either $a$ or $b$. Hence,
$E_a \cup E_b$ is a neighbourly set of same cardinality.

\section*{References}

[1] A.V. Aho, J.E. Hopcroft, and J.D. Ullman. {\em{The Design and
Analysis of Computer Algorithms}} Addison-Wesley Publishing Company,
1974.

[2] N.Alon, R.Yuster, and U.Zwick. Finding and counting given length
cycles. {\em{Algorithmica}}, 17:209--223, 1997.

[3] C.L.Barrett, V.S.A.Kumar, M.V.Marathe, S.Thite, and G.Istrate.  
Strong edge coloring for channel assignment in wireless radio networks.
In {\em{Proceedings of the 4th annual IEEE international conference on
Pervasive Computing and Communications Workshops}}, PERCOMW '06, pages
106--, 2006. 

[4] G.J.Bezem and J. van Leeuwen. Enumeration in graphs. Technical Report
RUU-CS-87-07, Department of Information and Computing Sciences, Utrecht
University, 1987.

[5] K.Cameron. Induced matchings. {\em{Discrete Applied Mathematics}},
24(1-3):97 -- 102, 1989.

[6] N.Chiba and T.Nishizeki. Arboricity and subgraph listing algorithms.
{\em{SIAM J. Comput.}}, 14(1):210--223, 1985. 

[7] A.Itai and M.Rodeh. Finding a minimum circuit in a graph. {\em{SIAM
Journal on Computing}}, 7(4):413--423, 1978.

[8] S.Isobe, X.Zhou, and T.Nishizeki. Total colorings of degenerate
graphs. {\em{Combinatorica}}, 27(2):167--182, March 2007. 

[9] T.R.Jensen and B.Tof. {\em{Graph coloring problems}}. Wiley
Interscience, 1995.

[10] S.Jukna. {\em{Extremal combinatorics - with applications in computer
science}}. Texts in theoretical computer science. Springer, 2001. 

[11] M.Mahdian. On the computational complexity of strong edge coloring.
{\em{Discrete Applied Mathematics}}, 118:239--248, 2002.

[12] A.Sen, M.Huson, A new model for scheduling packet radio
networks,{\em{ Proceedings of IEEE INFOCOM'96,}} Vol. 3, pp. 1116-1124,
1996.

\section*{Appendix}

Mahdian states and proves the following theorem:

{\textbf{Theorem 2:}[[11]]} In a quadrilateral-free graph $G$, for any
neighbourly set $A$ at least one of the following conditions hold:

(1) There exists a triangle $\triangle uvw$ such that $A \subseteq E_u
\cup E_v \cup E_w$

(2) There exists an edge $uv$ such that $A \subseteq E_u \cup E_v$

(3) There exists a vertex $u$ such that $A$ has at most $10$ edges not in
$E_u$

In [11] the author adds a remark, that the constant $10$ in Theorem 2 can
be replaced by $4$ and claims that would lead to an $O(n m^4)$ time
algorithm.  It appears that the claim in the remark is not correct as can
be seen from the following example (see Figure 19). 

\begin{figure}[h!] 
\centering
\includegraphics[width=2.3in]{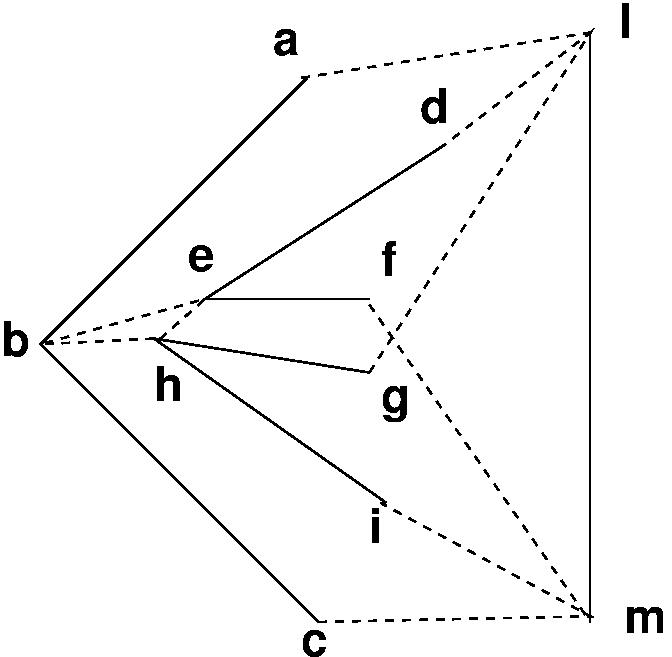}
\caption{Counter example for remark in [11]}
\end{figure}

Let neighbourly set $A$ consists of three disjoint paths $abc$,$def$ and
$ghi$ of length two (each) and an additional edge $lm$, vertex disjoint,
with these paths.  

The connecting non-neighbourly set edges in the graph are $be,bh,eh$ (as
in Figure 18) and $lm$ is connected with edges \\ $al, cm, dl, fm, gl,
im$\\ to each path of length two (as in Figure 15(b)). 

This graph is clearly quadrilateral free and $A$ is neither in form (1)
nor in form (2) of Theorem 2.

As each vertex has a maximum degree of two in $A$, and as there are seven
edges in $A$, there is no vertex $u$ for which $A$ has $4$ edges not in
$E_u$ (for example, five edges $de,ef,hg,hi,lm$ are not incident at $b$).

\end{document}